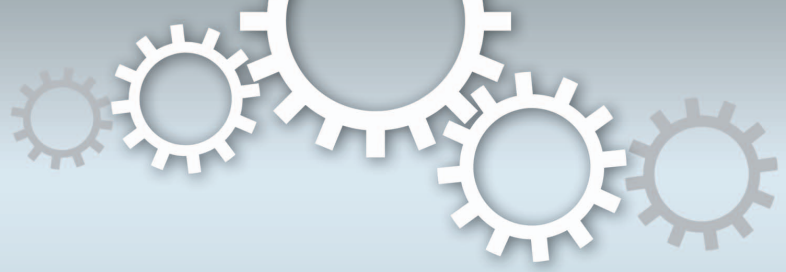

# SCIENTIFIC REPORTS

# Position-sensitive spectral splitting with a plasmonic nanowire on silicon chip

Qing Hu[1,2], Di-Hu Xu[1], Yu Zhou[1], Ru-Wen Peng[1], Ren-Hao Fan[1], Nicholas X. Fang[2], Qian-Jin Wang[1], Xian-Rong Huang[3] & Mu Wang[1]

[1]National Laboratory of Solid State Microstructures and Department of Physics, Nanjing University, Nanjing 210093, China, [2]Department of Mechanical Engineering, Massachusetts Institute of Technology, Cambridge, Massachusetts 02139, USA, [3]Advanced Photon Source, Argonne National Laboratory, Argonne, Illinois 60439, USA.

On-chip nanophotonics serves as the foundation for the new generation of information technology, but it is challenged by the diffraction limit of light. With the capabilities of confining light into (deep) subwavelength volumes, plasmonics makes it possible to dramatically miniaturize optical devices so as to integrate them into silicon chips. Here we demonstrate that by cascading nano-corrugation gratings with different periodicities on silver nanowires atop silicon, different colors can be spatially separated and chronologically released at different grating junctions. The released light frequency depends on the grating arrangement and corrugation periodicities. Hence the nanowire acts as a spectral splitter for sorting/demultiplexing photons at different nano-scale positions with a ten-femtosecond-level interval. Such nanowires can be constructed further into compact 2D networks or circuits. We believe that this study provides a new and promising approach for realizing spatiotemporal-sensitive spectral splitting and optical signal processing on nanoscales, and for general integration of nanophotonics with microelectronics.

S pectral splitting and imaging have numerous applications varying from optical communication, logical operations, micro-spectrum analyses, to photon sorting or sensing[1-6]. These spectral technologies have been based on traditional filters, mirrors, or interferometers, which are clumsy in size (much larger than the wavelength) and difficult to be integrated into microelectronics[7]. By relying on the silicon-on-insulator (SOI) technology[8-10], some on-chip spectral devices, such as light in/out couplers[11-13] and demultiplexers[14,15], have been successfully developed in the infrared regime, particularly for telecommunication wavelengths. Yet for visible frequencies, the silicon-based technology fails to work efficiently due to the strong inherent absorption of silicon. Consequently, on-chip integration of photonic devices has been limited by their larger-than-wavelength sizes and the weak optical response of silicon[16]. Fortunately, as a product of the interaction between photons and free electrons at the metal-dielectric interface, surface plasmons (SPs) can achieve extremely small mode wavelengths and high localized electromagnetic fields. Hence, plasmonics with (deep) subwavelength characteristics may break the diffraction limit of light, and thus are promising for modulating photons on nanoscales[17-19] so as to bridge the gap between nanoelectronics and optics[20-23]. In fact, plasmonic structures have shown potential applications as nano waveguides and circuits[24-26], nanoantennas[27,28], submicron dichroic splitters[29], photonic sorters[30], and logic gates[31], etc. However, to date it is still very challenging to make plasmonic devices compatible with the current complementary-metal-oxide-semiconductor (CMOS) technology, mainly due to the large impedance mismatch across the metal-dielectric interface[32,33]. Recently, by introducing hybrid plasmon polaritons[34,35], researchers have experimentally realized low-loss optical waveguiding on deep subwavelength scales[36], where the higher optical loss mode, termed as the dielectric-loaded-surface-plasmon-polariton (DLSPP) mode[37,38], is intentionally converted to a lower hybrid loss mode. But this technique again remains incompatible with the current silicon technology. Therefore, the key issue to advance this field is to explore the possibilities of integrating plasmonic nanostructures on silicon wafers, which is obviously of significant importance for combining nanophotonics and microelectronics and has been attracting intense interest in recent years.

Here we report the first experimental realization of a novel on-chip plasmonic device for position-sensitive spectral splitting of photons with cascading corrugation gratings along silver nanowires atop silicon wafers, where different colors are spatially separated and subsequently released at ten-femtosecond-level intervals. As shown in Fig. 1a, the fundamental mechanisms of this approach are as follows. On the microstructured silver nanowire, the input light is converted into broadband SPs, which then propagate along the wire and are modulated by the plasmonic band gaps of the periodic corrugation gratings. At the junction sites where the spatial periodicity of the corrugations is abruptly changed, SPs with a specific wavelength are selectively released. Thus, the structured silver nanowire acts as a position-sensitive spectral splitter (PSS) for photons, as well as a time-of-flight monitor[39] for femtosecond optical pulses. Such microstructured silver wires can also be constructed further into miniature 2D/3D networks/circuits for high-resolution spatiotemporal spectral splitting and imaging[40,41], optical signal demultiplexing on nanoscales, and many other potential applications of on-chip nanophotonics.

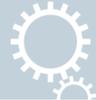

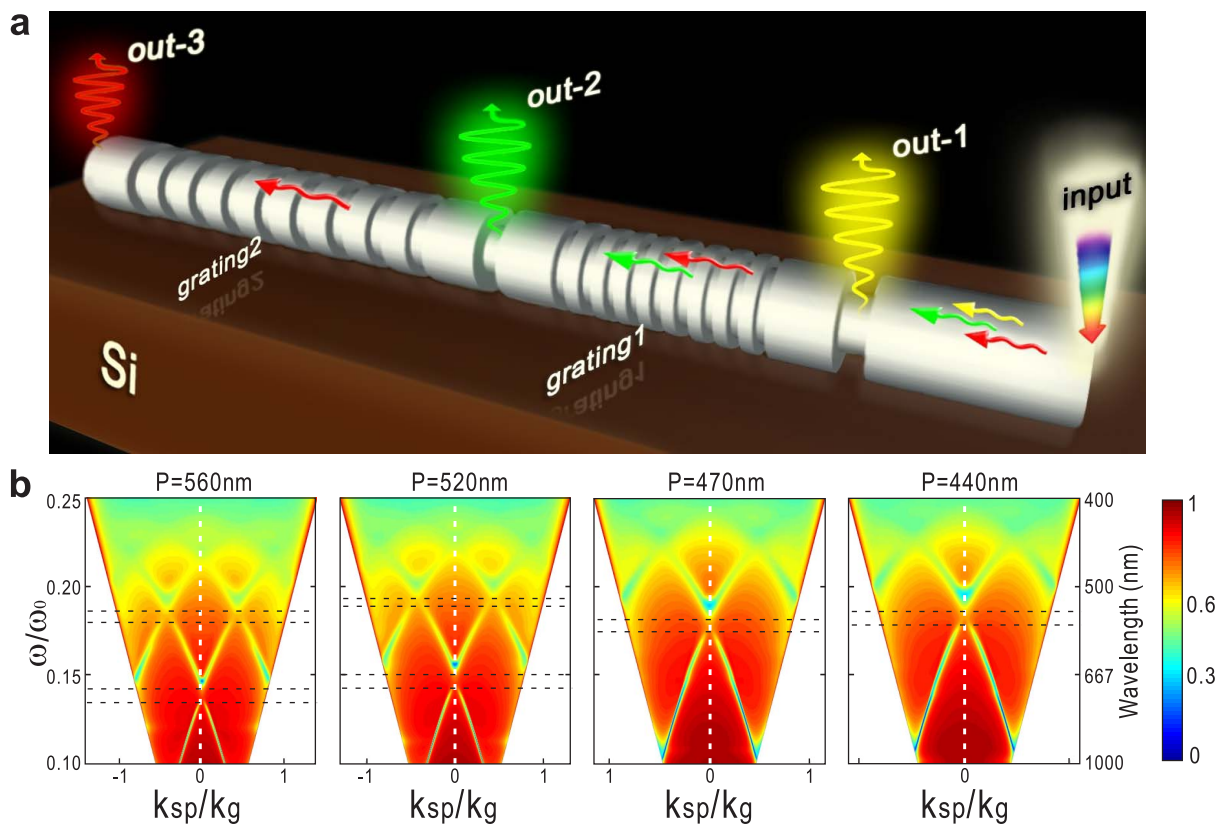

**Figure 1 | The principle of the plasmonic position-sensitive spectral splitter (PSS) device.** (a). Schematic of the plasmonic PSS consisting of two cascading corrugations with different spatial periodicities (marked as "grating1" and "grating2", respectively) for transporting and filtering SPs, and the grooves ("out-1" and "out-2") for exporting SPs. (b). Plasmonic band structures of four (individual) corrugations with periods $P = 560$ nm, 520 nm, 470 nm, and 440 nm, respectively. Here the frequency unit is $\omega_0 = 2\pi c/a$ ($c$ the speed of light in vacuum and $a = 100$ nm), the wave-vector unit is $k_g = 2\pi c/P$, and the color bar shows the reflection intensity. The blue-green arcs indicate the dispersion relation of the SPs in the structures. The band gaps appear at the edges of the Brillouin zones marked by the dashed lines. In both the calculations and experiments, the nanowire diameter is always fixed at 170 nm, and the width and depth of the corrugations are both set at 30 nm.

## Results

**Concept of a plasmonic nanowire-based spectral splitter atop silicon chip.** A smooth silver nanowire without corrugations on a dielectric surface usually supports several types of SP modes, such as guiding modes and leaking modes[42,43] with a continuous dispersion relation in a wide frequency range[44,45] (Supplementary Fig. S1a). Thus, it is believed that the silver nanowire on a high-permittivity wafer (silicon, for example) cannot transport regular SP modes[46]. However, a broadband SP mode, which is a hybrid gap mode trapped between the nanowire and the wafer, can be supported (see Supplementary Information). Yet this hybrid SP mode has a small mode area corresponding to large propagation loss, which is still unfavorable for large-distance propagation. To solve this problem, here we intentionally coat an ultrathin $SiO_2$ layer on the silver nanowire and achieve a lower hybrid loss mode[36]. This type of hybrid mode possesses much smaller mode areas ($0.06A_0$, $A_0 = \lambda^2/4$) and larger propagation lengths (around 15 μm at $\lambda = 647$ nm). Consequently, the converted broadband SP modes are able to propagate along the nanowire with much less loss.

Furthermore, by introducing a grating consisting of periodic corrugations on the silver nanowire (shown in Supplementary Fig. S1b), these SP modes can be modulated by the plasmonic band structure that is determined by the spatial periodicity of the grating. Figure 1b shows the plasmonic band structures of four individual gratings with periodicities of $P = 560$ nm, 520 nm, 470 nm, and 440 nm, respectively. It is obvious that the band gaps appear at $n\frac{\pi}{P}$ (Brillouin zone edges, $n$ being an integer), where the SPs are strongly back-reflected and cannot propagate forward[23]. In our approach, cascading grating segments with different periodicities are constructed on the silver wire, as schematically illustrated in Fig. 1a. Between the cascading grating segments a single groove is introduced, where the localized SPs in the band gap, which possess specific colors, are released back into light by scattering of this groove. In this configuration, the single grooves act as "bus stops" for different colors. Therefore, the periodic corrugation gratings marked as "grating 1" and "grating 2" play the role to transport and filter SPs, while the grooves "out-1" and "out-2" act as emitters to release the filtered SPs. Overall, when a beam with mixed wavelengths is introduced into this structured nanowire, broadband SP modes are excited and propagate along the nanowire. Thus, "grating 1" transmits the SPs with frequencies falling within its propagation bands while prohibiting those SPs with frequencies falling in the band gap. Then the prohibited SPs are released at "out-1". Similarly, the SPs that pass through "grating1" but are prohibited by "grating 2" are released at "out-2". The SPs passing through both "grating 1" and "grating 2" are released by "out-3", and so on. With this mechanism, the functionality of subwavelength position-sensitive spectral splitting, demultiplexer or light in/out coupling is realized.

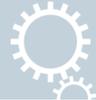

**Experiments for a parallel plasmonic PSS.** Firstly, we have fabricated a structured silver nanowire (diameter 170 nm) containing two corrugation gratings with periodicities $P = 470$ nm ("grating 1") and $P = 520$ nm ("grating 2"), respectively, as shown in Figs. 2a and 2b. In the middle is a single groove separating the two gratings. The input laser beam is focused on this single groove. The SPs excited by this groove propagate towards the two ends of the nanowire. As mentioned above, since the plasmonic band gaps of the two gratings are different, the output colors at the two ends can be different.

The optical measurements have been carried out with the experimental setup schematically illustrated in Supplementary Fig. S2. The light signals detected at the middle groove and the two ends for different input wavelengths are shown in Figs. 2d–2g. When the wavelength of the input laser is $\lambda = 647$ nm, both terminals show red bright spots in Fig. 2d. This is reasonable since the wavelength is within the propagation bands of two corrugations with $P = 470$ nm and $P = 520$ nm, as indicated in Fig. 1b. When the input wavelength is changed to $\lambda = 568$ nm (Fig. 2e), there is no output from "out-1" (dark) whereas the end "out-2" turns yellow. This indicates that for this wavelength, propagation of the SPs is prohibited in "gratings 1" but allowed in "grating 2", which is again consistent with Fig. 1b. When the input wavelength is further changed to 530 nm, "out-1" turns green and "out-2" becomes dark in Fig. 2f since SPs corresponding to $\lambda = 530$ nm are forbidden in "grating 2". The whole optical spectra of the scattered light measured at the two ends are shown in Figs. 2c (middle and bottom). Compared with the spectrum of a smooth silver nanowire without corrugations (Fig. 2c (top)), the intensity for $\lambda = 568$ nm drops dramatically at "out-1", whereas the intensities for $\lambda = 514$, 520 and 530 nm attenuate obviously at "out-2". These experimental data are in good agreement with the calculated normalized transmission (the red lines in Fig. 2c for continuous wavelength variations).

It should be pointed out that in our experiments, the output intensity of light in the propagation wavelength bands of the corrugation gratings, such as 480–540 nm and 630–700 nm in Fig. 2c (middle), maintains the same order of magnitude as that for the smooth nanowire (Fig. 2c (top)). This shows that the transmission efficiency of the SPs on the corrugated structure is high. For example, comparing with the same-length silver nanowire without corrugations, the light going through "grating 2" has the output efficiency around 75% at $\lambda = 647$ nm, and 85% at $\lambda = 568$ nm, respectively.

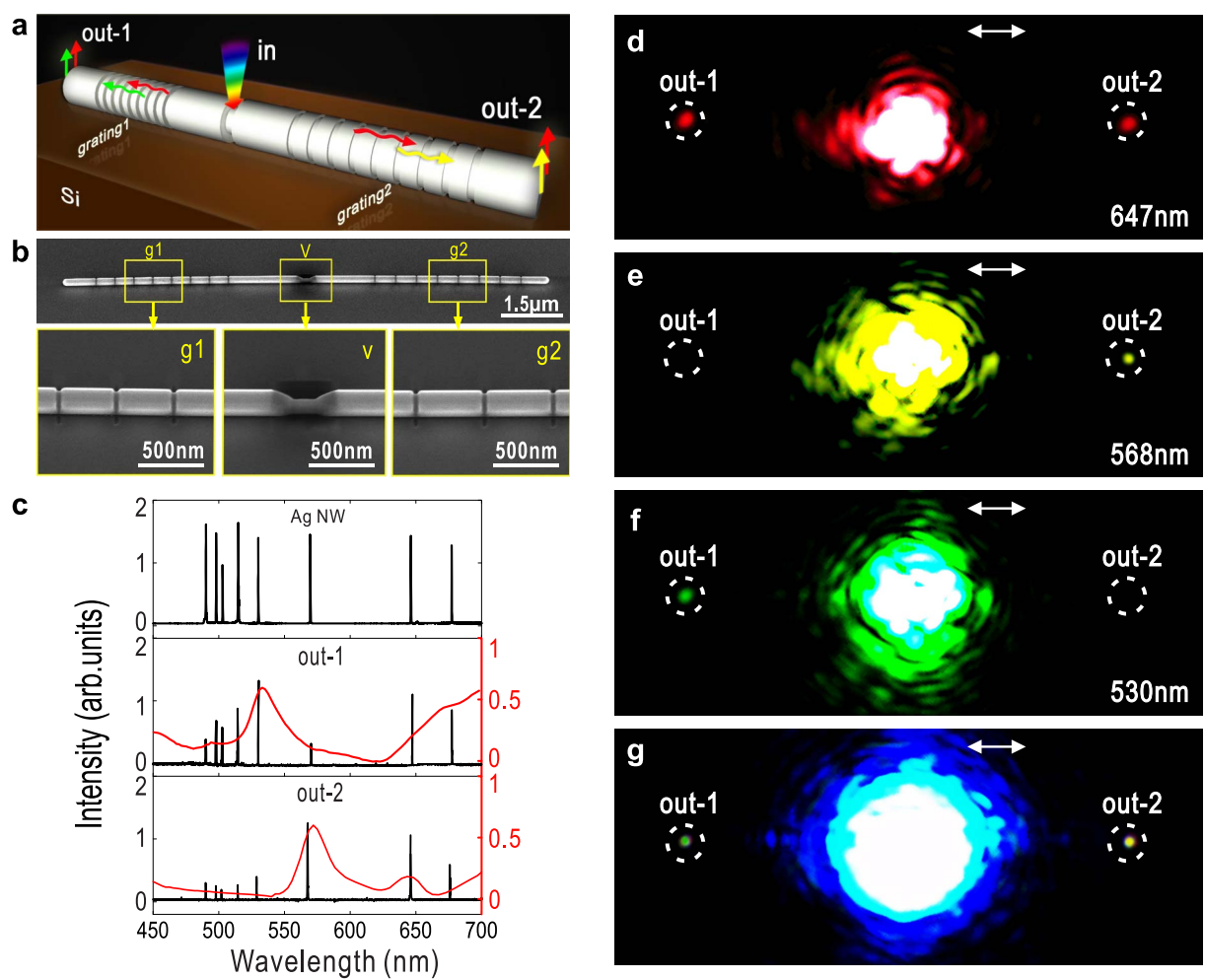

**Figure 2 | Transportation of modulated SPs on the corrugated silver nanowire.** (a). Schematic of the structured silver nanowire with two corrugated gratings (with periods $P = 470$ nm and $P = 520$ nm) separated by a single groove in the middle that converts the incident beam into SPs propagating towards the two ends of the nanowire. (b). SEM images of the nanowire and the enlarged structure details. The nanowire diameter is about 170 nm; the corrugation period is about 470 nm for "grating 1", and 520 nm for "grating 2", respectively. (c). **Top:** Scattered light spectra measured from the end of a smooth nanowire with no corrugations. **Middle and bottom:** Experimental (black lines) and calculated (red lines) scattered light spectra from "out-1" and "out-2" of the corrugated nanowire, respectively. (d–f). Emission images of "out-1", the middle groove, and "out-2" for incident wavelengths of 647 nm, 568 nm and 530 nm, respectively. (g). The corresponding emission images for a polychromatic input laser beam, where both ends have output but with different selected colors. The white double-arrow (parallel to the nanowire) indicates the polarization direction of the input beam.

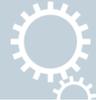

**Experiments for a cascade plasmonic PSS.** Now we focus on the propagation of light along the silver nanowire with cascading corrugation gratings in Figs. 3a (also see the schematic of Fig. 1a). This nanowire (170 nm in diameter) possesses two sets of periodic corrugations, one with periodicity $P = 470$ nm ("grating 1") and the other with $P = 520$ nm ("grating 2"). The two grooves corresponding to "out-1" and "out-2" in Fig. 1a are clearly shown in the SEM images of Fig. 3a (here "out-3" is simply the left end of the nanowire). When a polychromatic laser beam is focused at the input terminal (right most in Fig. 3a), SPs are generated and propagate along the nanowire towards the gratings. When passing through "grating 1", the SP corresponding to $\lambda = 568$ nm is forbidden to propagate, so it is scattered at "out-1", resulting in a yellow bright spot at "out-1" in Fig. 3b. The rest of SPs continue to propagate along the nanowire. When they reach "grating 2", the SPs with $\lambda = 480$–540 nm are forbidden to propagate. Consequently, they are released at "out-2" as a green bright spot. The remaining SPs are finally exported to the terminal "out-3", where they are released as a red bright spot in Fig. 3b. Figures 3c–e are the optical spectra detected at the three output positions, from which one may identify the strong peaks corresponding to the released SPs in Fig. 3b. These measured spectra agree quite reasonably with the calculated spectra of the red curves in Figs. 3c–e. Some minor discrepancies are mainly due to the fact that the chemically-synthesized nanowire usually does not have strictly round cross-sections or perfectly smooth surface. In fact, the cross-sectional shape of the nanowire indeed affects the confinement of light[47], i.e., different cross-sectional shapes may change the hybrid optical mode along the wire. More detailed consideration for the discrepancies between experiments and simulations are presented in Supplementary Information. In our experiments, a number of silver nanowire samples with different diameters have been fabricated and optically measured. For instance, we have also made the 260 nm-diameter nanowire with cascade periodic corrugations of $P = 440$ nm ("grating 1") and $P = 520$ nm ("grating 2") (see Supplementary Information), similar good agreement between experiments and simulations has been achieved. With these data, we conclude that the position-sensitive spectral splitting mechanism has been indeed realized by corrugated silver nanowires on silicon wafers.

It is interesting to characterize further the performance of the on-chip plasmonic PSS working at visible frequencies. The spectral selectivity of the splitter is one of the key parameters for practical applications, which can be described by the side-band suppression[48,49] of the plasmonic structure. Here the side-band suppression is defined as $S_{dB} = -10 \log_{10} [I(\lambda)/I_0(\lambda)]$, where $I(\lambda)$ is the wavelength-dependent intensity collected from the nanowire with corrugated gratings, and $I_0(\lambda)$ is the collected intensity from a similar nanowire but without corrugations. For example, the above 170 nm-diameter corrugated silver nanowire (Fig. 3a) possesses the side-band suppression $S_{dB} \equiv 34$ dB at $\lambda = 568$ nm, which may be further improved by changing the diameter of the wire (see the data in Table I). Besides, it is also worthwhile to evaluate the out-coupling efficiency of the released light from the PSS, which can be characterized by $\eta(\lambda) = P(\lambda)/P_0(\lambda)$, where $P_0(\lambda)$ is the wavelength-dependent power of incident light propagating along the nanowire, and $P(\lambda)$ is the power of the released light. According to Fig. 3c, the outcoupling efficiency of the yellow light ($\lambda = 568$ nm) can reach about 47% at "out-1" in the 170 nm-diameter device; and it can be further enhanced to about 74% when the wire diameter is increased to 260 nm. Yet the outcoupling efficiency of green light ($\lambda = 520$ nm) at "out-2" dramatically attenuates due to the intrinsic loss of SPs. Anyway, based on Table I, one can still expect that within a limited working distance (e.g. 15 μm), high outcoupling efficiency and large side-band suppression can be achieved simultaneously in the plasmonic PSS designs by optimizing the geometry and dimensions of the nanowires. In this way, the performance of the plasmonic

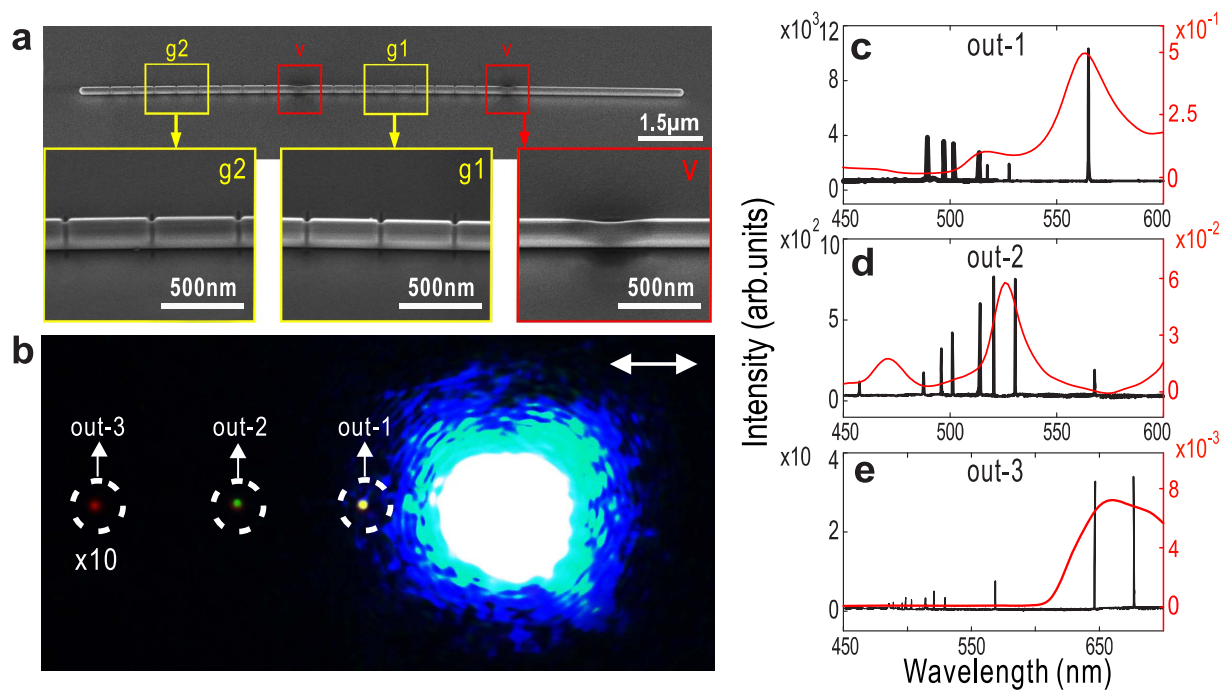

**Figure 3 | Selective propagation and emission of SPs along a silver nanowire with cascading corrugation gratings.** (a). SEM images of the silver nanowire with cascading gratings (corresponding to Fig. 1a) and the enlarged structure details: corrugations with $P = 520$ nm (left: grating 2), corrugations with $P = 470$ nm (middle: grating 1), and the "bus-stop" groove (right: out-1). The nanowire diameter is about 170 nm. (b). Emission micrograph of the structured nanowire illuminated by a polychromatic laser beam from the input end (rightmost end). The bright spots marked by the dash white circles indicate the light scattered from grooves "out-1", "out-2" and "out-3" (leftmost end). The white double-arrow indicates the polarization direction of the input beam. (c–e). Experimental (black lines) and calculated (red lines) scattered light spectra from "out-1", "out-2" and "out-3", respectively.

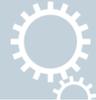

| Table I | The performance of the cascade PSS based on two types of silver structured nanowires with different diameter (D) on silicon chip | | | |
|---|---|---|---|---|
| Beam via the wire | $\lambda = 568$ nm | | $\lambda = 520$ nm | |
| Performance | $D = 170$ nm | $D = 260$ nm | $D = 170$ nm | $D = 260$ nm |
| Side-band suppression | 34 dB | 28 dB | 31 dB | 26 dB |
| Outcoupling efficiency | 47% @ ''out-1'' | 74% @ ''out-1'' | 6% @ ''out-2'' | 17% @ ''out-2'' |
| Mode area ($A_0$) | 0.07 | 0.17 | 0.06 | 0.12 |

PSS at visible frequencies can match that of non-plasmonic grating couplers (such as SOI waveguides) at infrared wavelengths[11–13,50–52]. On the other hand, with the help of the hybrid SP modes, the plasmonic PSS possesses much smaller mode areas and thus achieves much better confinement of light compared to non-plasmonic devices. For instance, the 170 nm-diameter structured silver nanowire in Fig. 3a has the mode area $A_{pss} \cong 0.06A_0$ (with $A_0 = \lambda^2/4$), whereas the non-plasmonic waveguides usually have the mode area $A_{dw} \geq A_0/n^2$ ($n$ is the refractive index of the dielectric waveguide material). Given that silicon has high loss at visible frequencies, non-plasmonic waveguides are usually based on other dielectric materials with relatively lower index $n$. Thus for the confinement of visible light, the plasmonic PSS works ten times better than non-plasmonic devices.

**Time-domain analysis of a cascade PSS.** To understand the transportation process of the SPs in the PSS, we have carried out numerical simulations using the finite-difference time-domain (FDTD) method. Figure 4a shows the calculated electric field distributions around the corrugated nanowire in Fig. 3a for incidence wavelengths of 568 nm, 530 nm and 647 nm, respectively. Apparently, the excited SPs are selectively guided along the nanowire. Around "out-1" in Fig. 4a, the forward propagation of SPs corresponding to $\lambda = 568$ nm is blocked, hence the SPs are scattered to the far field.

SPs corresponding to $\lambda = 530$ nm pass through "grating 1" but are blocked around "out-2". In contrast, SPs for $\lambda = 647$ nm propagate through both "grating 1" and "grating 2". These FDTD simulations again clearly verify the above results, i.e., different colors are spatially separated along the nanowire due to the microstructures. It should be noted that different colors are released chronologically at 15-femtosecond-or-so intervals, as shown in Figs. 4b–d. Therefore, the structured nanowire simultaneously act as a time-of-flight monitor for femtosecond optical pulses in addition to its function of nanoscale position-sensitive spectral splitting.

## Discussion

It is worthy to emphasize that the present plasmonic PSS possesses the following characteristics. First, the plasmonic nanowire with corrugations can realize on-chip spectral splitting in the visible frequency band. For example, when white light passes through the cascade PSS in Fig. 3a, the outcoupling efficiency of yellow light ($\lambda = 568$ nm) can reach 47%, and meanwhile the side-band suppression can reach about 34 dB. By further optimization of the wire geometry and dimension, the spectral splitting performance of the plasmonic PSS at visible frequencies can match that of non-plasmonic grating couplers (such as SOI waveguides) at infrared wavelengths[51,52]. For comparison, SOI waveguides fail to work efficiently at visible frequencies because of the strong inherent absorption of silicon. Second, the plasmonic PSS (based on a 170 nm-diameter silver wire, for example) can achieve ten-times more efficient optical confinement because it possess a much smaller mode area ($0.06A_0$, $A_0 = \lambda^2/4$) comparing with the non-plasmonic waveguides for visible wavelengths. Third and more importantly, the plasmonic PSS can be fully compatible with current CMOS techniques, which makes it possible to develop on-chip photonic devices and to integrate nano-photonics with microelectronics.

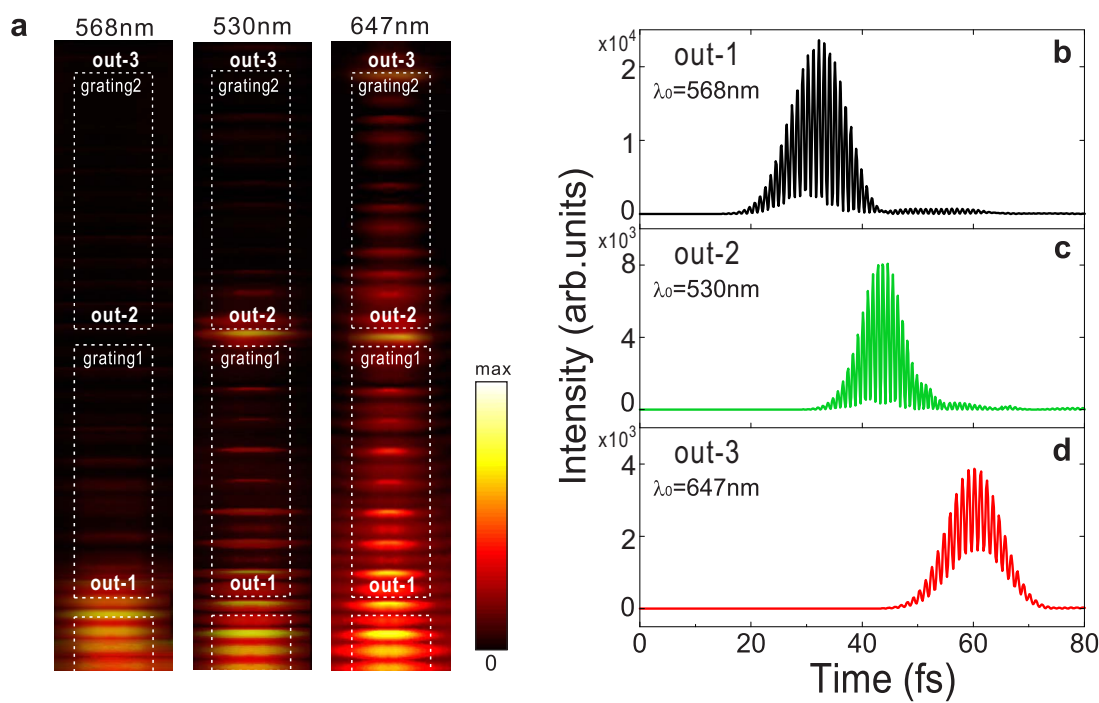

**Figure 4 | The FDTD simulation of the transport process of SPs on the structured nanowire of Fig. 3.** (a). The calculated electric field distributions ($|\mathbf{E}|^2$) along the nanowire (diameter 170 nm) for incidence wavelengths of 568 nm (left), 530 nm (middle) and 647 nm (right), respectively. The white dash boxes indicate the corrugation gratings. The gap between neighboring dashed boxes corresponds to a "bus stop" groove. (b–d) The calculated temporal spectra of the output light pulses for wavelengths 568 nm, 530 nm and 647 nm, respectively. Different colors are released chronologically at 15-femtosecond-or-so intervals. In the simulation the incident laser pulse is assumed to be a focused Gaussian beam with a pulse length of 2 fs and a central wavelength 568 nm, 530 nm and 647 nm, respectively.

Furthermore, based on corrugated silver nanowires on silicon chip, we can construct plasmonic networks to perform two-dimensional (2D) position-sensitive spectral splitting. Figure 5a schematically illustrates such a cross network, which consists of four branches of silver nanowires (170 nm diameter) with different corrugation periodicities of 520 nm, 560 nm, 470 nm, and 440 nm, respectively.

When a white beam is incident on the center of the cross, the electric-field distributions ($|E|^2$) of the SPs propagating in the network have been numerically calculated at wavelengths 568 nm (yellow), 530 nm (green), and 488 nm (blue), and 647 nm (red), respectively. The calculations demonstrate that the yellow beam propagates to the "North" (Fig. 5b), the blue beam propagates to the "East" (Fig. 5c), the green beam propagates to both the "West" and "South" (Fig. 5d), and the red beam can reach all directions (Fig. 5e). With the 2D scheme, therefore, parallel demultiplexing and processing of light signals can be realized. In addition, the pixels can be flexibly tuned in two dimensions for spectral splitting and imaging.

In summary, by fabricating corrugated silver nanowires on silicon wafer, we have realized for the first time nanoscale position-sensitive spectral splitting of photons as well as time-of-flight monitoring of femtosecond optical pulses. Moreover, the corrugated nanowires can be constructed into compact 2D spatiotemporal-sensitive networks. Our studies provide a new approach to develop spatiotemporal-resolved spectral splitting networks on nanoscales. Such devices may also be applied for optical frequency division and optical signal demultiplexing on nanoscales. Most significantly, the plasmonic devices are ultracompact on silicon wafers and the nanofabrication process is fully compatible with current CMOS techniques, which demonstrates the feasibility of design and fabrication of on-chip nanophotonics combined with microelectronics. The development of spatiotemporal optical signal processing on the nanoscales together with other recent achievements (e.g., nanophotonic logics and networks[31,53,54]) could make plasmonic circuits more reality for the next generation information technology in the future.

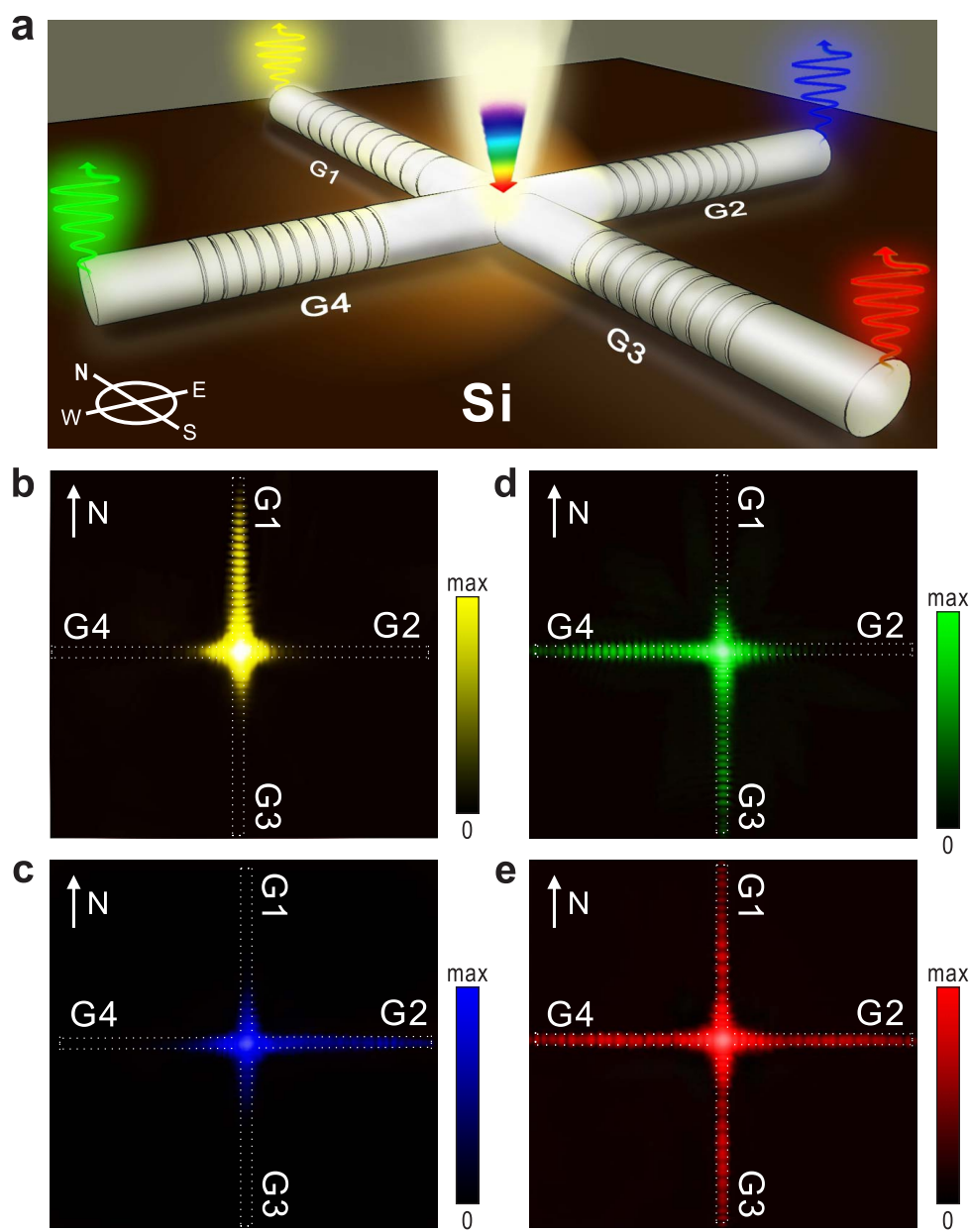

**Figure 5 | The FDTD simulation of a 2D PSS with structured cross nanowire networks.** (a). Schematic of the 2D network consisting of four silver nanowire arms with different corrugation periods: 520 nm (G1), 560 nm (G2), 470 nm (G3) and 440 nm (G4), respectively. The wire diameter is always fixed at 170 nm, and the width and depth of the corrugations are both set at 30 nm. Each grating consists of eight corrugations, starting from the location 3 μm away from the central cross. (b–e). The calculated electric-field distributions ($|E|^2$) of the SPs propagating in the network with wavelengths 568 nm, 488 nm, 530 nm, and 647 nm, respectively. The arrow "N" indicates the north direction. In the simulation, the incident light source is an unpolarized Gaussian beam focused on the central cross. When a white light incidents at the center cross, SPs with specific colors propagate along different channels.

## Methods

**Sample fabrication.** The silver nanowires are chemically synthesized[55] with diameter 170 nm and length around 15 μm. The wires are then transferred onto a polished silicon substrate, followed by a coating of 5 nm-thick layer of $SiO_2$ via magnetron sputtering. The corrugations are fabricated on the silver nanowires by focus-ion-beam milling (FIB, Helios Nanolab 600i). Both the width and the depth of the corrugation are fixed at 30 nm, and the spatial periodicity varies from 400 nm to 600 nm. The profile of the "bus stop" groove between different sections of corrugations is optimized in order to increase the coupling efficiency between the incident light and SPs.

**Experimental setup.** As schematically shown in Supplementary Fig. S2, the input polychromatic laser beam (with main wavelengths 647 nm, 568 nm, 530 nm, 520 nm, 514 nm, 488 nm from Spectra-Physics Lasers, 2018-RM) is focused by an objective (100×) to the end or the middle of the nanowire. The emitted light from the wires is analyzed by a spectrometer (Princeton Instruments,SP-2500). The emission is also imaged by a CCD. The spectrum measurement and the emission imaging are switchable by a mirror.

**Simulation.** Based on the full-wave finite-difference time-domain (FDTD) method[56], we carry out the numerical simulation on the transportation of SPs and the scattered light spectra on the structured silver nanowires by using a commercial software package (Lumerical, FDTD_Solutions version 8.0.1). The geometry is modeled as that of the measured samples. The simulation is modeled with optimum parameters of focused Gaussian sources and detector positions[57].

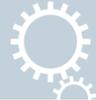

www.nature.com/**scientificreports**

## Acknowledgements
This work was supported by the Ministry of Science and Technology of China (Grant Nos. 2012CB921502 and 2010CB630705), the National Science Foundation of China (Grant Nos. 11034005, 61077023, and 11021403), and partly by the Ministry of Education of China (20100091110029). XRH was supported by the U.S. Department of Energy, Office of Science, Office of Basic Energy Sciences, under Contract No. DE-AC02-06CH11357. NXF acknowledges partial support by NSF Grant no. CMMI-1120724.

## Author contributions
R.W.P. and Q.H. developed the concept. Q.H., R.H.F., D.H.X. and Q.J.W. fabricated the samples. Q.H. characterized the samples. Q.H. and D.H.X. performed the measurements. Q.H. and Y.Z. performed the calculations. Q.H., R.W.P., N.X.F., M.W. and X.R.H. analyzed and interpreted the data and implications. Q.H., R.W.P., X.R.H. and M.W. wrote the manuscript. R.W.P. and M.W. supervised the researches.

## Additional information
**Supplementary information** accompanies this paper at http://www.nature.com/scientificreports

**Competing financial interests:** The authors declare no competing financial interests.

**How to cite this article:** Hu, Q. *et al.* Position-sensitive spectral splitting with a plasmonic nanowire on silicon chip. *Sci. Rep.* **3**, 3095; DOI:10.1038/srep03095 (2013).

# Supplementary Information

# Position-sensitive spectral splitting with a plasmonic nanowire on silicon chip

Qing Hu[1,2], Di-Hu Xu[1], Yu Zhou[1], Ru-Wen Peng[1,]*, Ren-Hao Fan[1], Nicholas X. Fang[2],

Qian-Jin Wang[1], Xian-Rong Huang[3], and Mu Wang[1,]*

*1) National Laboratory of Solid State Microstructures and Department of Physics,*

*Nanjing University, Nanjing 210093, China*

*2) Department of Mechanical Engineering, Massachusetts Institute of Technology, Cambridge,*

*Massachusetts 02139, USA*

*3) Advanced Photon Source, Argonne National Laboratory, Argonne, Illinois 60439, USA*

**Outline**

**Supplementary Figures**

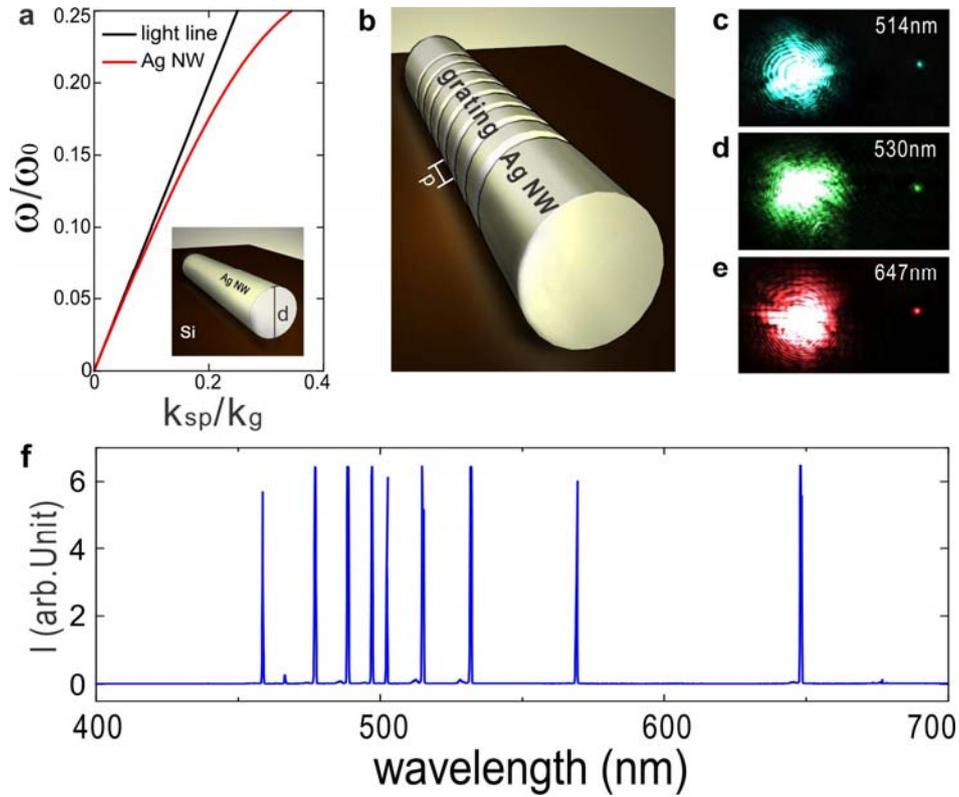

**Supplementary Figure S1 | SP propagation on Ag nanowire atop Si wafer. a.** Dispersion relation of a smooth non-corrugated silver nanowire on a Si substrate showing a wide continuous frequency range. Inset: schematic of silver nanowire with radius (d=170nm) on Si wafer. Microscopic images taken by CCD for illuminated laser source with wavelength as **c.** 514nm; **d.** 530nm; **e.** 647nm, respectively. **f.** Measured output spectra of a silver nanowire with length around 8μm under multiple lasers illumination with the wavelengths as 488nm, 514nm, 520nm, 530nm, 568nm, 647nm.

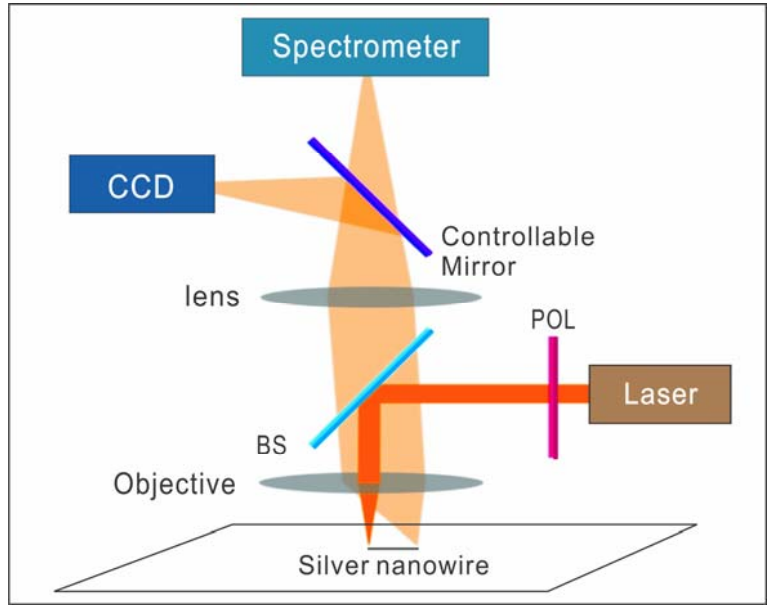

**Supplementary Figure S2 | Experimental setups.** POL: polarizer; BS: beam splitter. The input laser beam (with main wavelengths as 647 nm, 568 nm, 530 nm, 520 nm, 514 nm, 488 nm from Spectra-Physics Lasers, 2018-RM) is focused by an objective (100x) to the end or the middle of the nanowire. The emitted light is analyzed by a spectrometer (Princeton Instruments，SP-2500). The emission from the microstructured nanowire is imaged by a CCD. The spectrum measurement and the emission imaging are switchable by a controllable mirror.

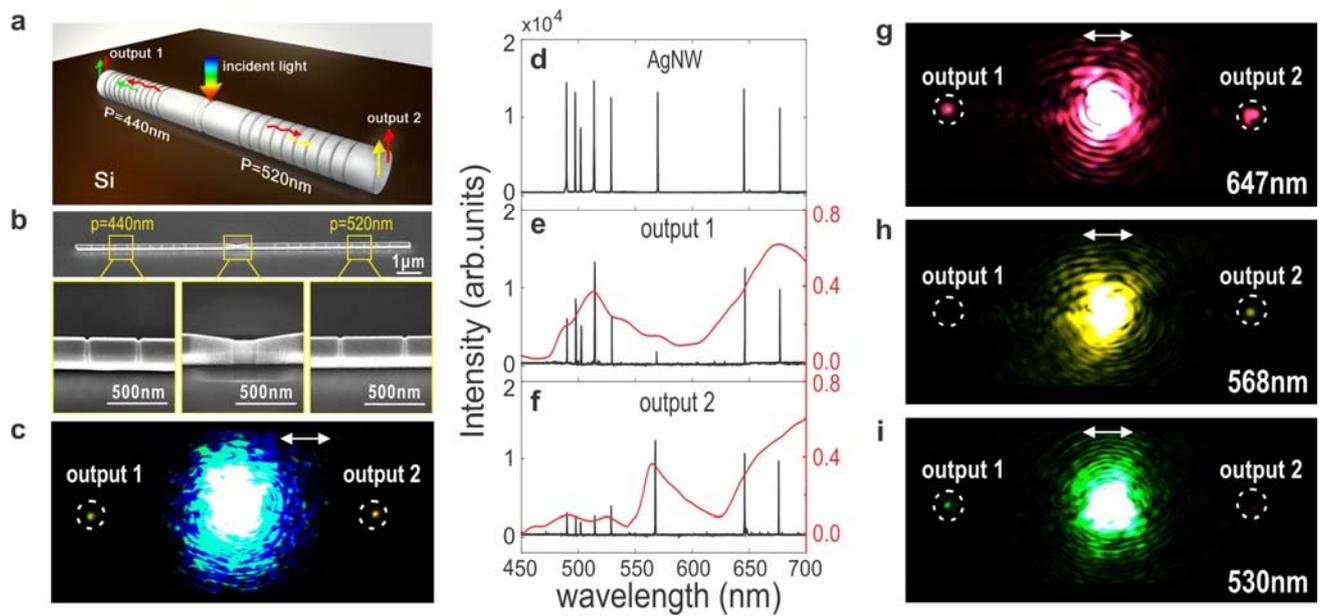

**Supplementary Figure S3 | The additional experiment for parallel plasmonic PSS based on a 260nm-diameter Ag nanowire consisting of two corrugated gratings with periodicity as 440nm and 520nm, respectively. a.** Schematic of the structured silver nanowire with two corrugated gratings (with periods P = 440 nm and P = 520 nm) separated by a single groove in the middle that converts the incident beam into SPs propagating towards the two ends of the nanowire. **b.** SEM images of the nanowire and the enlarged structure details. **c.** The corresponding emission images for a multiple-wavelength laser beam. **d.** Scattered light spectra measured from the end of a smooth nanowire with no corrugations. **e, f.** Experimental (black lines) and calculated (red lines) spectra taken from "output1" and "output2" of the corrugated nanowire, respectively. **g-i.** Emission images of "output1", the middle groove, and "output2" for incident wavelengths of 647nm, 568nm and 530nm, respectively.

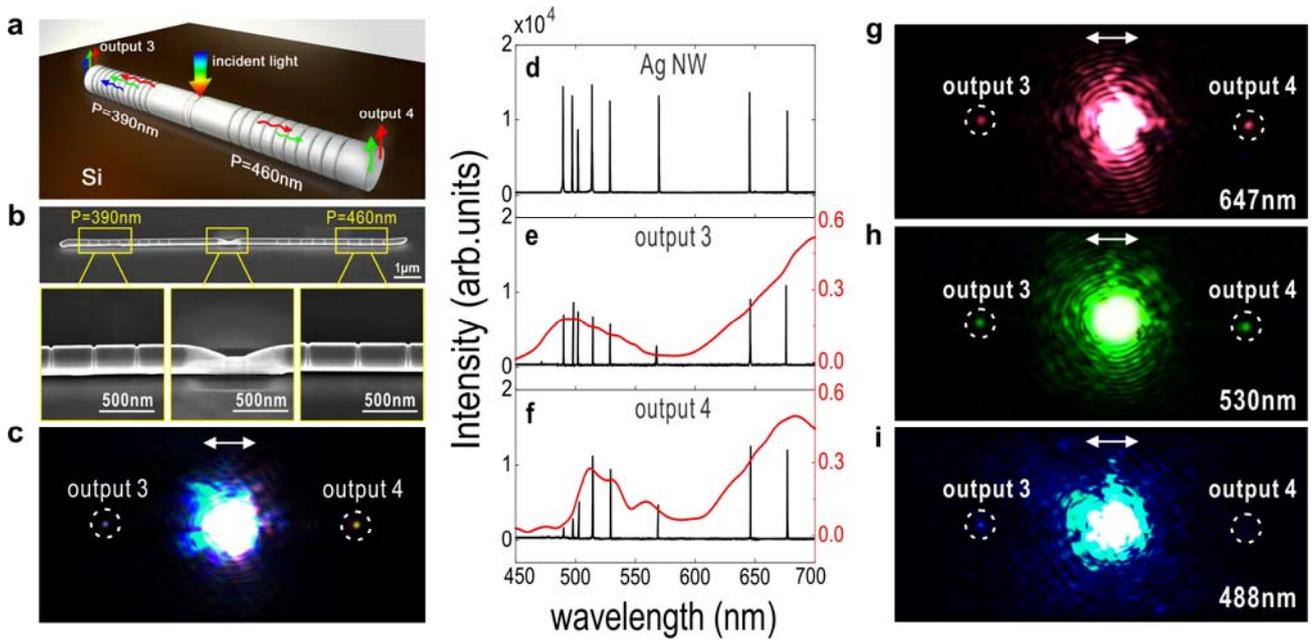

**Supplementary Figure S4 | The additional experiment for parallel plasmonic PSS based on a 260nm-diameter Ag nanowire consisting of two corrugated gratings with periodicity as 390nm and 460nm, respectively. a.** Schematic of the structured silver nanowire with two corrugated gratings (with periods P = 390 nm and P = 460 nm) separated by a single groove in the middle that converts the incident beam into SPs propagating towards the two ends of the nanowire. **b.** SEM images of the nanowire and the enlarged structure details. **c.** The corresponding emission images for a multiple-wavelength laser beam. **d.** Scattered light spectra measured from the end of a smooth nanowire with no corrugations. **e, f.** Experimental (black lines) and calculated (red lines) spectra taken from "output3" and "output4" of the corrugated nanowire, respectively. **g-i.** Emission images of "output3", the middle groove, and "output4" for incident wavelengths of 647nm, 530nm and 488nm, respectively. The white double-arrow segment in **c, g-i** indicates the polarization direction of the input beam.

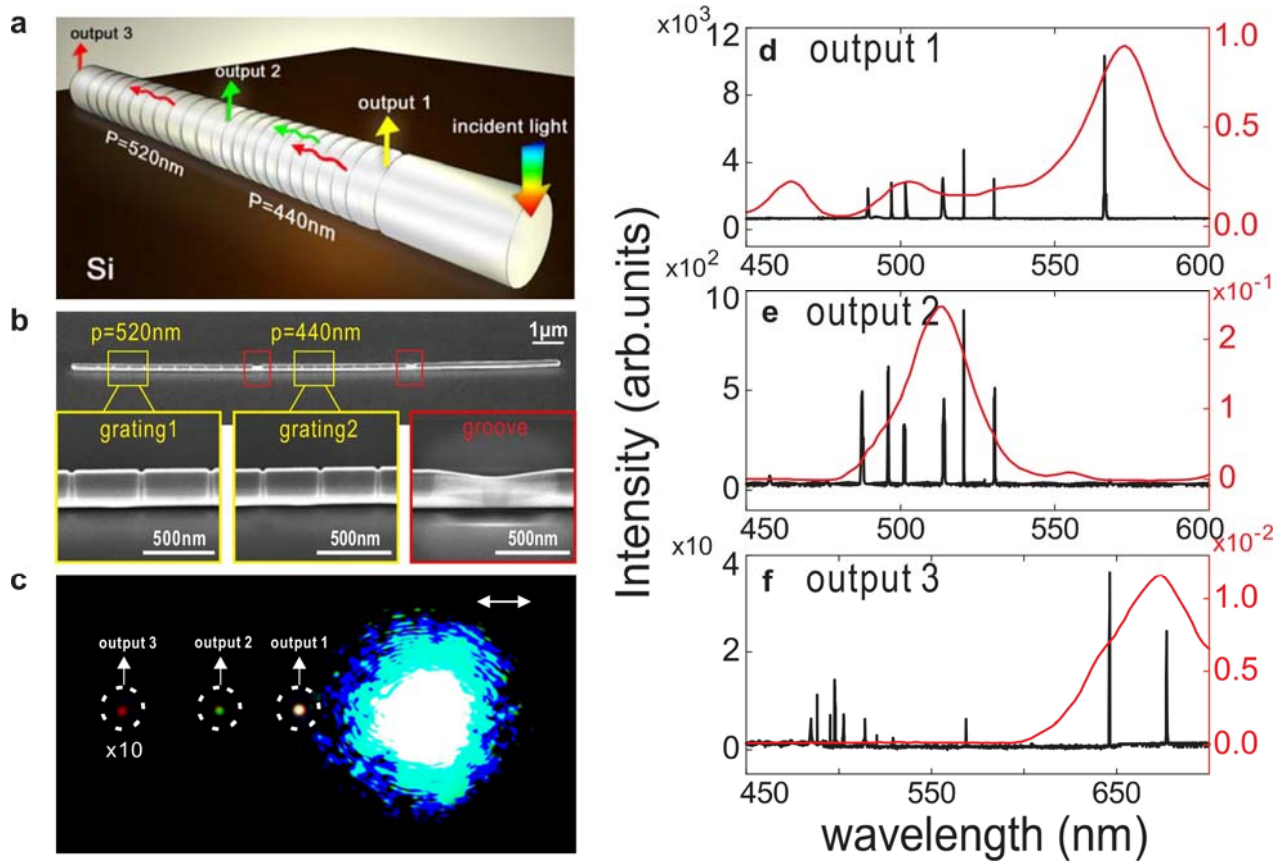

**Supplementary Figure S5 | The additional experiment for cascaded plasmonic PSS based on a 260nm-diameter Ag nanowire consisting of two corrugated gratings with periodicity as 440nm and 520nm, respectively). a.** Schematic illustration of a silver nanowire with cascading corrugation with period P=440nm (termed as Grating 2) and P=520nm (termed as Grating 1), respectively. Two grooves are marked as "Output 1" and "Output 2", respectively. **b.** SEM images of the configuration of the structures: corrugations with P=520nm (left), corrugations with P=440nm (middle), and one groove (right). **c.** Emission micrograph of the structured nanowire illuminated by a multiple-wavelength laser beam from the input end (right most). The bright spots marked by the dash white circles indicate the light scattered from grooves as "output1", "output2" and "output3". The white double-arrow segment indicates the polarization direction of the input beam. **d-f.** Experimental (black lines) and calculated (red lines) scattered light spectra taken from sites "output1", "output2" and "output3", respectively.

**Supplementary Calculations and Analysis**

**1. Surface plasmon modes on Ag nanowire atop Si wafer**

The dispersion relation of the SP mode on smooth Ag nanowire on Si wafer can be calculated by using Finite Element Method (Comsol 3.5a). As shown in Fig. S1a, the dispersion is continuous in a wide frequency region, which means the Ag nanowire supports a broadband SP mode. In the experiments, we deposit the chemically synthesized Ag nanowires [S1] (with radius around 85nm) on the surface of Si wafer. When focusing laser beams at one distal facet, the surface plasmons can be launched to propagate along the nanowire axis, then couple to far field light at wire discontinuities [S2]. The scattering images at different propagation wavelengths (514nm, 530nm, 647nm) are detected with a charge-couple-device camera (CCD) camera (Fig. S1c-e), respectively. The spectra are measured by a spectrometer under illuminations of multiple lasers with main wavelengths as 488nm, 514nm, 520nm, 530nm, 568nm, 647nm (Fig. S1f). These figures demonstrate that the Ag nanowires on Si wafer are able to support SP modes at a wide frequency range.

To evaluate the propagating properties of the SPs, the effective mode index ($n_{eff}$) and propagating length ($L$) are calculated by using finite element method. Here the effective index and propagating length are defined as $n_{eff} = \beta/k_0$ and $L = 1/2\operatorname{Im}\{\beta\}$, respectively, where $\beta$ is the propagation constant of the SP mode and $k_0$ is the wave vector in vacuum, respectively. As shown in Fig.S6a, with the refractive index of the substrate increasing, $n_{eff}$ increases, while $L$ decreases. It indicates that the higher-refractive-index substrate results in higher loss, and the hybrid gap mode has higher loss than the axis-symmetric mode. The mode profiles in the three situations are shown in Fig. S6b-d. It can be seen that due to the high-refractive index and optical absorption of Si, most of the electric energy is confined in the gap between Ag nanowire and Si substrate (Fig. S6d). When decreasing refractive index of substrate, more electric energy lies in the substrate (Fig.S6c). Further decreasing the refractive index until there is no difference of the refractive index between surrounding media and substrate material leads to an axis-symmetric SP mode (Fig.S6 b). These results indicate that the nanowire sustains a hybrid gap mode of SPs at the presence of substrate. The profile of the hybrid mode shows great dependence on the refractive index of substrates. The similar phenomenon has been reported in Ref. S3-S5.

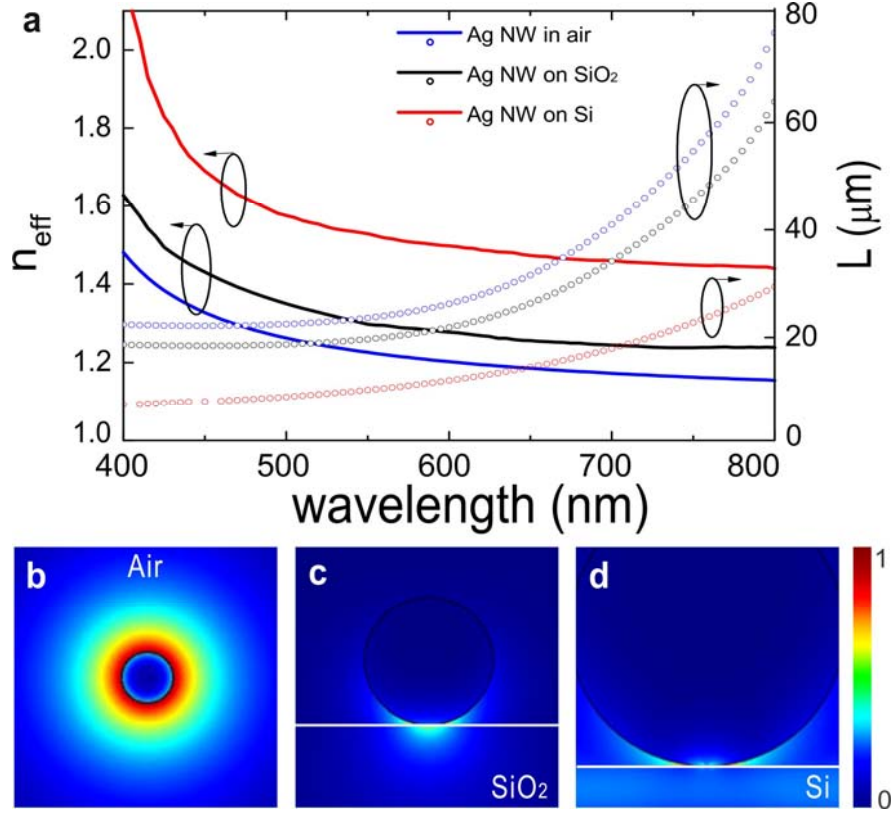

**Supplementary Figure S6 | Propagation behavior of SP mode on Ag nanowire. a.** Effective Mode index and propagation length of SP modes sustained by Ag nanowire with a diameter as 170nm in air, on SiO$_2$ substrate, and on Si substrate, respectively. Mode profiles of the SP modes (647nm) sustained by Ag nanowire **b** in air. **c** on SiO$_2$ substrate. **d** on Si substrate (enlarged shows the electric energy in the gap), respectively.

In addition to the effective mode index and propagation loss, the effective mode area $A_m$ has also been investigated. The effective mode area $A_m$ is defined as the ratio of mode energy per unit length along the direction of propagation and its peak energy density, such that [S6]

$$A_m = \frac{W_m}{\max\left\{W(r)\right\}} = \frac{1}{\max\left\{W(r)\right\}} \int_{-\infty}^{\infty} W(r) d^2r \tag{1}$$

where $W_m$ and $W(r)$ are the electromagnetic energy and energy density (per unit length along the direction of propagation), respectively:

$$W(r) = \frac{1}{2}\left(\frac{d\left(\varepsilon(r)\omega\right)}{d\omega}\left|E(r)\right|^2 + \mu_0 \left|H(r)\right|^2\right) \tag{2}$$

We calculate the normalized ode area $A$ (defined as $A_m/A_0$, where $A_0$ is the diffraction-limited area in free space with the value as $\lambda^2/4$) of Ag nanowire with different sizes (170nm and 260nm radius, respectively) at visible region. Here the Ag nanowire is coated with 5nm SiO$_2$, positioned on top of Si substrate, which is the same as the experimental situation. As shown in Fig. S7a, the mode area of 260nm Ag nanowire is lower than 0.2. When decreasing the diameter of Ag nanowire to 170nm, the mode area is smaller. At some wavelength points, e.g.568nm, it can reach 0.02. Therefore the confinement could be enhanced by scaling down the size of nanowire. From the mode profiles (Fig. S7b,c), we can see that the electric energy is both confined around the surface of SiO$_2$ coated Ag nanowire and in the gap between the nanowire and the substrate. Thus the modes are hybrid SP modes.

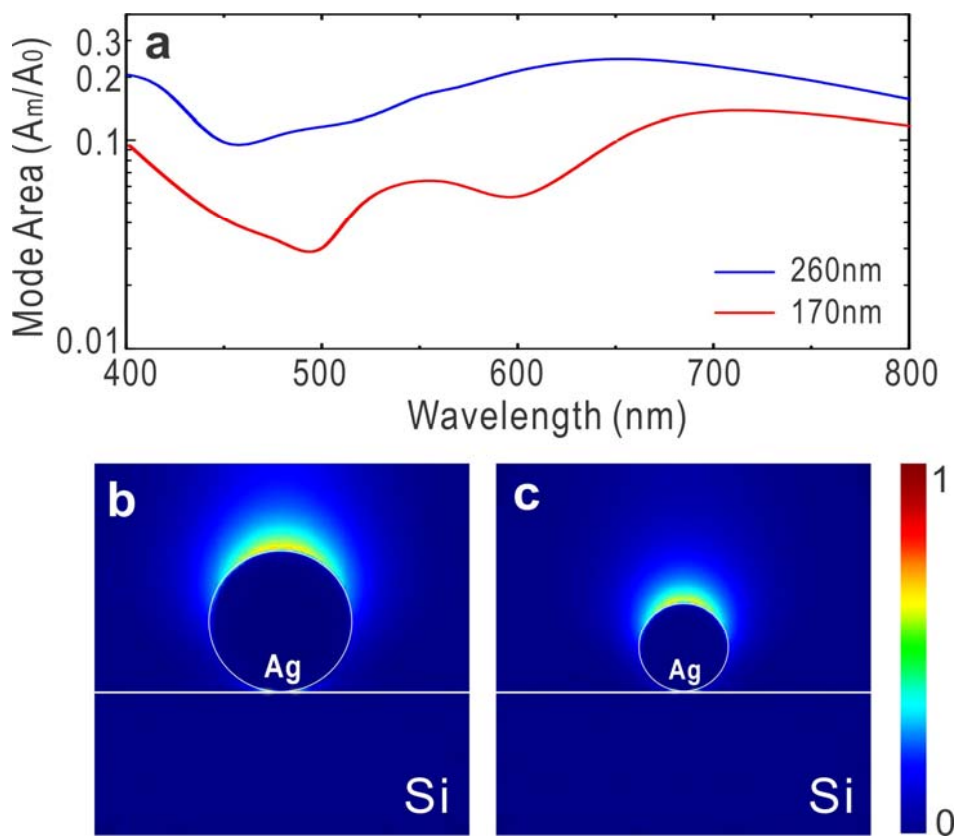

**Supplementary Figure S7 | Effective mode area and mode profile of SPs on Ag nanowire with 5nm SiO$_2$ coating layer. a** Effective mode areas of SPs on Ag nanowire with different diameters (blue line: 260nm, red line: 170nm) atop Si substrate. Here the values of mode areas are plotted on a logarithmic scale. **b,c** Mode profiles at the wavelength as 647nm for 260nm-diameter and 170nm-diameter nanowires, respectively. The data is calculated by using FDTD method (Lumerical 8.0.1).

## 2. Plasmonic band gaps of corrugated Ag nanowire atop Si wafer

For a suspended metallic cylindrical waveguide whose diameter is reduced to several hundreds of nanometer, only the fundamental (azimuthally symmetric) transverse magnetic SP mode exists [S7]. According to the solutions of Maxwell's equations in cylindrical coordinate and the boundary conditions of SP wave, the dispersion of a metallic nanowire ($\varepsilon_1$, $R_1$) coated a dielectric layer ($\varepsilon_2$, $R_2$) embedded in a medium ($\varepsilon_3$) can be derived as

$$\varepsilon_1 k_1 I_1(k_1 R_1)\left[\varepsilon_3 k_2 K_1(k_3 R_2)M_{00} + \varepsilon_2 k_3 K_0(k_3 R_2)M_{10}\right] = -\varepsilon_2 k_1 I_0(k_1 R_1)\left[\varepsilon_3 k_2 K_1(k_3 R_2)M_{01} + \varepsilon_2 k_3 K_0(k_3 R_2)M_{11}\right]$$

(3)

where

$$M_{ab} = I_a(k_2 R_2)K_b(k_2 R_1) - (-1)^{a+b} I_b(k_2 R_1)K_a(k_2 R_2)$$

(4)

$k_j = (\beta^2 - \varepsilon_j k^2)^{1/2}$, $I_a$ and $K_a$ are the modified Bessel functions of the first and second kinds, respectively. The calculated dispersions of the SP mode are shown in Fig.S8a. It can be seen that SP mode of Ag nanowire with larger radius approaches to the mode on an Ag-air surface, which indicates a similar behavior of the SP mode between the metallic nanowire and planar cases. Therefore the metallic nanowire can be approximately treated as a planar metallic surface.

For a metallic nanowire atop a silicon wafer, by optimizing the permittivity of dielectric ($\varepsilon$=2.7 in our case), the dispersion of a planar Ag-dielectric surface can be fitted to that of a 5nm-SiO$_2$ coating Ag nanowire (Fig. S8b). Thus the coated Ag nanowire is reasonable to be treated as a planar Ag-dielectric surface. Therefore it is reasonable to apply the planar Ag/dielectric model to generate the calculations results in SiO$_2$-coating Ag nanowire structure. Under the approximation situation, by using the rigorous coupled-wave analysis (RCWA) method [S8], angular-dependent reflection spectra can be generated to describe the plasmonic band gaps. In the reflection spectra (as shown in Fig 1b), the minimum reflectivity corresponds to the photons that have been absorbed through the excitation of SPs. Therefore the green-blue arcs in the figures denote the dispersions of the SP modes on the textured Ag nanowire. Due to the periodically corrugations, photonic gaps appear at the Brillouin zone edges, where the SPs are back-reflected so strongly that they cannot propagate any more. With increasing the periodicity of the corrugations, the central frequencies of the plasmonic gaps have red shifts. Besides, one more band gap appears at high frequency region.

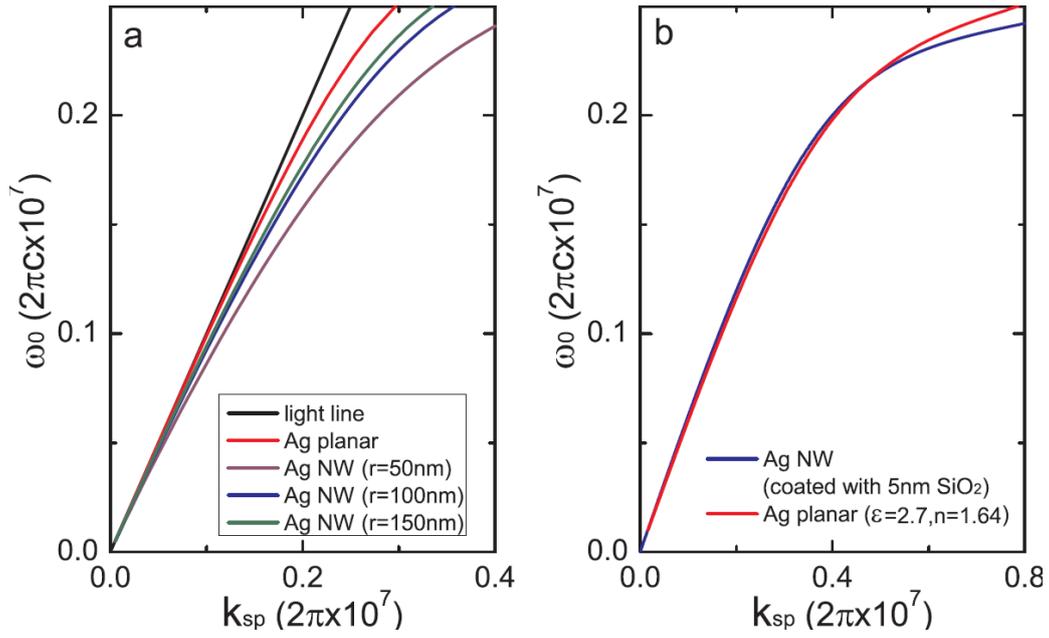

**Supplementary Figure S8 | Dispersion relations of SP modes on Ag nanowires and Ag planar. a** Dispersion relations of SP modes on Ag nanowires with different radiuses. **b** Dispersion relation of a planar Ag-dielectric film (the permittivity of dielectric is set as 2.7) agrees well with that of a 5nm-SiO$_2$-coated Ag nanowire.

**Supplementary Experiments and Discussions**

**1.  Experimental setup and measurement**

We measure the spectra by using a confocal micro spectrometer system. As schematically shown in Fig. S2, the input laser beam (Spectra-Physics Lasers, 2018-RM with main wavelengths as 647nm, 568nm, 530nm, 520nm, 514nm, 488nm) is focused on the input positions (the distal ends or the grooves in the middle ) by an objective (100x). The emitted spectra are measured by a spectrometer (Princeton Instruments，SP-2500). The scattered image is detected by a CCD. The spectrum measurement and the emission imaging are switchable by a controllable mirror. In the measurement, we fix the laser power, tilting the beam splitter (shown in Fig. S2) to focus the incident laser on the input terminals. Slightly tune the objective focusing to find largest scattered intensities, then record the spectra.

**2.  Additional experimental demonstrations**

To confirm the functions of the plasmonic PSS, we fabricate the similar structures on 130nm-radius silver nanowires. We firstly fabricated the parallel PSS, in which two corrugations with periodicities P=440nm and P=520nm are separated by a single groove (Fig. S3a,b). When focusing the laser beam on the groove, the SPs are excited to propagate towards the two ends of the nanowire. Due to the manipulation of plasmonic band gaps of the two gratings, the SPs on the silver nanowire are selected to propagate, thus the output colors at the two ends are different (Fig. S3g-i). The optical spectra of the scattered light are measured at the two ends (Fig. S3e,f). Compared with the spectra of a smooth silver nanowire (Fig. S3d), the intensity for λ=568nm drops at "output1" in Fig S3e, while the intensities for λ=514, 520, 530nm attenuate obviously at "output2" in Fig. S3f. These measured data are in good agreement with the calculated normalized transmission (red lines in Fig. S3e,f). We further fabricate another parallel PSS with different grating periodicities as 390nm and 460nm. As shown in Fig. S4, the propagation of SPs are tuned by the plasmonic band gap, and different frequencies are selected to output at the two ends. Then we fabricate the cascaded PSS. As shown in Fig. S5a-b, the structure consists of two gratings with periods as P=440nm (grating 2) and P=520nm (grating 1), respectively, and two grooves termed as "output 1" and "output 2", respectively. Because the plasmonic band gaps of the gratings forbid some frequencies and let the others go through, different colors of the released light from the grooves. From the scattering image detected by CCD (Fig. S5c), we observe a yellow bright spot at "output1" and a green bright spot at "output2". Figure S5d-f show the optical spectra measured by the spectrometer (black lines), together with the calculated ones (red curves). These measured and calculated spectra are in agreement with each other. With these data, we confirm that the cascaded PSS can be realized based on the structured silver nanowire. It should be noted that because the difference of the effective mode index between the Ag nanowires with radius as 85nm and 130nm is small, the gratings on the two types of Ag nanowires have nearly the same periods (520nm & 470nm for 85nm-radius nanowire and 520nm & 440nm for 130nm-radius nanowire).

## 3. Side-band suppression and outcoupling efficiency

We perform FDTD simulation (Lumerical 8.0.1) to get the side-band suppression data. In the simulation, the scattering energy are collected from the output positions of two kinds of nanowires without grating coupler ($I_0(\lambda)$) and with grating coupler ($I(\lambda)$). Then the side-band suppression is calculated as

$$S_{dB}(\lambda) = -10\log_{10}[I(\lambda)/I_0(\lambda)] \tag{1}$$

This parameter evaluates the ability of the grating to forbid the SPs in the band gaps. Therefore it can be used to judge the quality of the plasmonic PSS. In our structures, the 170nm-diameter PSS reaches the side-band suppression $S_{dB}$ =34dB at λ=568nm, and $S_{dB}$ =31dB at λ=520nm, and the 260nm-diameter PSS reaches the side-band suppression $S_{dB}$ =28dB at λ=568nm, and $S_{dB}$ =26dB at λ=520nm (data shown in Table **I**), which can be compared with that of non-plasmonic grating coupler [S9, S10]. From Table **I**, it is interesting to see that PSS with smaller diameter has better side-band suppression than a larger one. It owes to better confinement of a smaller nanowire, which makes the manipulation of the grating more efficient. Thereafter the side-band suppression can be improved by changing the size of the nanowire.

In our case, the desired output light beams are designed out of the nanowire, and hence the outcoupling efficiency of each released light frequency is defined as

$$\eta(\lambda) = \frac{P(\lambda)}{P_0(\lambda)}, \tag{2}$$

where $P(\lambda)$ is the light power out of the grating, $P_0(\lambda)$ is the incident surface wave power propagating along the nanowire. Thus the value can be evaluated by the data in Fig. 3c, d and Fig. S5d,e. The outcoupling efficiency of 568nm light at "out-1" reaches about 47% in the 170nm-diameter PSS, while dramatically increases to 74% in the 260nm-diameter PSS. Similarly, the outcoupling efficiency of 520nm light at "out-2" is about 6% in the 170nm-diameter PSS, while reaches to 17% in the 260nm-diameter PSS. From the data we can have two conclusions. Firstly, the outcoupling efficiency of the 260nm-diameter PSS is much larger than that of the 170nm-diameter. It is resulted from the effect of the size of nanowire on the SPs loss. Usually small structure has better confinement but with large loss. Secondly, the outcoupling efficiency dramatically attenuates from "out-1" to "out-2" due to the intrinsic loss of SPs.

## 4. Discrepancy between experiments and simulations

It can be seen from Fig. 3, 4 and Fig. S3, S4 although the experimental data agree with calculated ones, there are still discrepancies between experiments and simulations. The discrepancy, on one hand is due to the fact that the chemically-synthesized nanowire usually cannot generate nanowires with perfectly rounded cross-sectional shapes. Recently it has been studied that the cross-sectional shape of the nanowires will affect SPs propagation behavior [S11]; on the other hand, is resulted from the defects on the Ag nanowires (Fig. S9), which couldn't be taken account of in the calculations. Besides, due to the limitation of fabrication, the scratches on Si wafer cannot be avoided when milling the grooves on Ag nanowire by using FIB. These scratches also introduce mismatch between the experimental and calculated results.

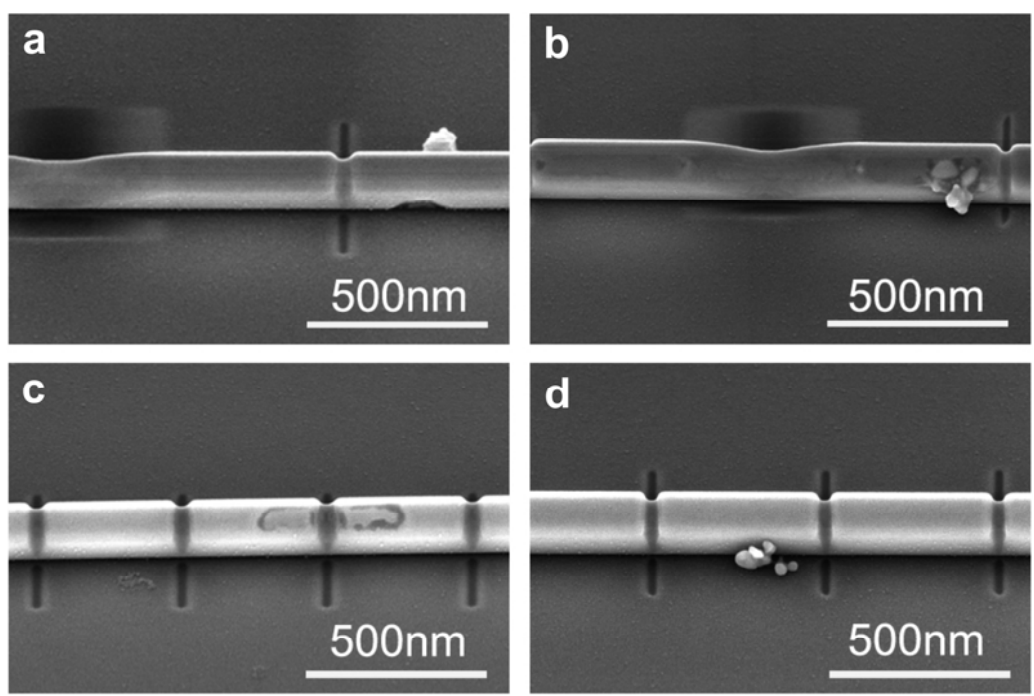

**Supplementary Figure S9 | SEM images of Ag nanowires with defects.**

## 5. Propagation performances of the plasmonic PSS and SOI nanowire at visible regime

Silicon-on-insulator (SOI) has received much attention and gain great achievement in the recent few years as it offers the possibility of integrating optical devices into electronic circuits by using the existing Si miscorelectrics fabrication technology. Currently, SOI waveguide based devices which operate at infrared wavelength of 1.3μm have already been integrated on submicro-size chip. Next we investigate the performances of the plasmonic PSS and SOI nanowire at visible regime by comparing the propagation length and mode confinement. Considering the top-down fabrication technology, the SOI nanowire may be easy to be fabricated to have a rectangle cross section. In the comparison, the Ag nanowire has a 170nm diameter and 5nm $SiO_2$ coating layer, which is exactly same as our experimental situation. In the SOI nanowire structure, a 50nm-thick $SiO_2$ spacer layer is positioned between the Si nanowire and Si substrate. The Si nanowire atop the $SiO_2$ has a square cross-sectional shape, whose side length equals the diameter of Ag nanowire. We calculate the propagation length and mode profile by using FDTD method (Lumerical 8.0.1) with a Gaussian source at 647nm operating wavelength. As shown in Fig. S10a, even within a 3μm distance the transportation energy in SOI nanowire is lower than 0.2 comparing with the input power. Hence the SOI nanowire fails to propagation the light at 647nm wavelength. The electric energy is confined around the Ag nanowire surface (Fig. S10b), while distributed in the whole Si nanowire area (Fig. S10d). These differences reveal the advantages of on-chip surface plasmonic devices at visible regime.

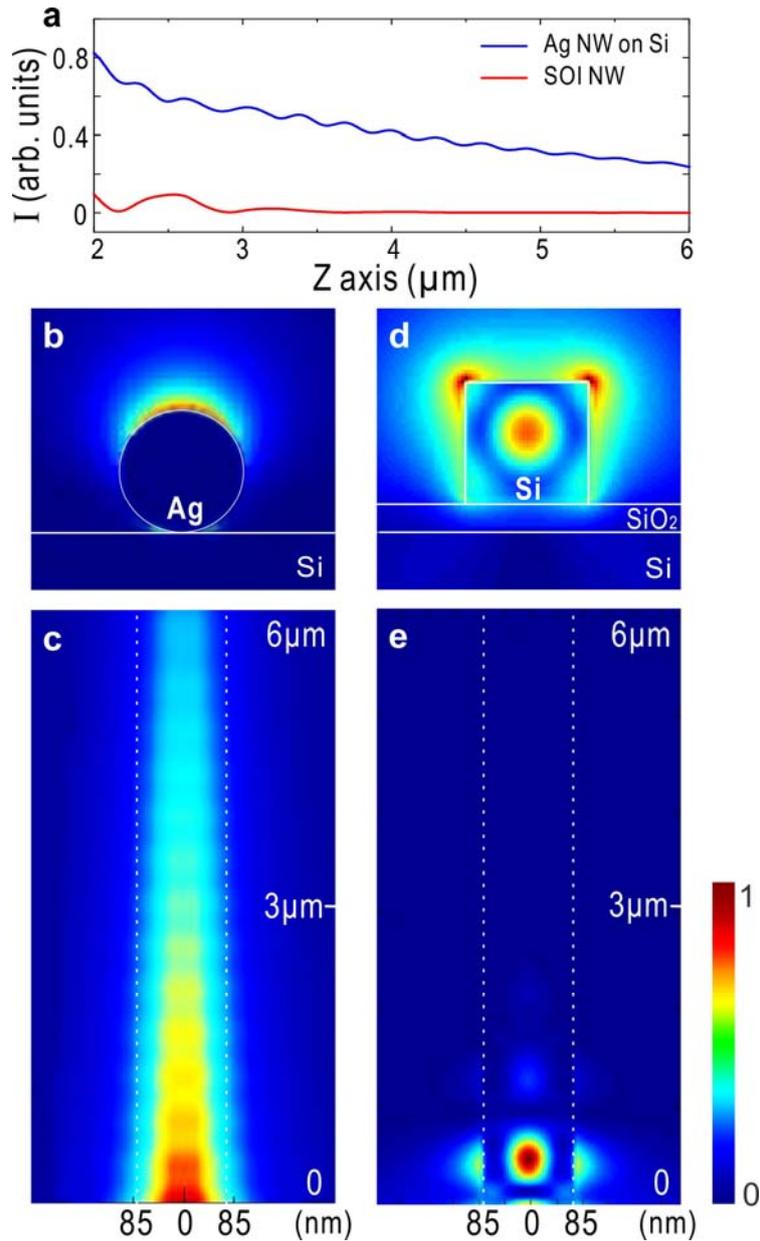

**Supplementary Figure S10 | Propagation performance of the plasmonic PSS and SOI nanowire at visible regime. a.** Normalized intensity ($|\mathbf{E}|^2$) collected from the output distal end of Ag nanowire (blue line) and SOI nanowire (red line) with different lengths. Mode profiles of **b.** Ag nanowire, **d.** SOI nanowire. Normalized intensity ($|\mathbf{E}|^2$) detected above a 6μm length **c.** Ag nanowire, **e.** Si nanowire. The observation plane of Ag nanowire is located at 10nm from the wire top, while the observation plane of Si nanowire is located at the center of the wire. In the SOI nanowire structure, the thickness of the spacer layer (SiO$_2$) between the Si nanowire and Si substrate is set as 50nm. Both the diameter of Ag nanowire and side width of Si nanowire are 170nm.